\documentclass[letterpaper,prb,twocolumn,showpacs,superscriptaddress]{revtex4} \usepackage{graphicx} \usepackage{amsmath}
\begin{document}
\title{Fluctuations in the relaxation dynamics of mixed chaotic systems}
\author{Roy Ceder and Oded Agam} \affiliation{The Racah Institute of Physics, The Hebrew University, Jerusalem 91904, Israel} \date{\today}
\begin{abstract}
The relaxation dynamics in mixed chaotic systems are believed to decay algebraically with a universal decay exponent that emerges from the hierarchical structure of the phase space. Numerical studies, however, yield a variety of values for this exponent. In order to reconcile these result we consider an ensemble of mixed chaotic systems approximated by rate equations,
and analyze the fluctuations in the distribution of Poincar\'{e} recurrence times. Our analysis shows that the behavior of these fluctuations, as function of time, implies a very slow convergence of the decay exponent of the relaxation. \end{abstract} 
\maketitle
\section{Introduction}
Recurrences play an important role in physics. The statistics of the recurrence times of particles govern the transport properties of open systems such as quantum dots attached to external leads, and the relaxation characteristics of distribution functions in closed systems, see Ref.~[\onlinecite{MeissReview}]. Additionally, in the short wavelength limit, quantum mechanical properties such as energy level correlations\cite{AgamAndreevAltshuler}, weak localization\cite{AleinerLarkin} and shot noise\cite{AgamAleinerLarkin}, are also determined by recurrences of the underlying classical dynamics.

One of the basic statistical characteristics of recurrences in open systems is the distribution of Poincar\'{e} recurrence times, $F(t)$. This is the probability of a trajectory to return, at time larger than $t$, to a predefined region in phase space. In this paper we focus our attention on the  the distribution of Poincar\'{e} recurrence times of "mixed" chaotic systems (more precisely-one dimensional symplectic maps). In these generic systems the phase space consists of islands of regular dynamics immersed in a sea of chaotic behavior, and around these islands there are smaller satellite regular motion islands around which are even smaller islands, and so forth {\it ad infinitum}\cite{MeissReview}. In such systems, a trajectory within the chaotic component of the phase space may stick to the regular islands for an exceedingly long time. This feature is believed to manifest itself in an algebraic decay of the distribution of Poincar\'{e} recurrences in the long time asymptotic limit:
\begin{equation}
F(t) \sim t^{-\gamma_0}. \label{power-law}
\end{equation}

Many studies have investigated this algebraic decay, focusing on two main issues: The question of universality of the decay exponent $\gamma_0$, and the calculation of its actual value. Yet, after three decades of studies, the answers to these questions are still controversial. Theoretical and numerical studies \cite{Channon80,Chirkov81,Chirkov85,MeissOtt,Grassberger85, Geisel87,Lai92,Shlesinger93,Zaslavsky97,Chirikov99,Weiss03,Cristadoro2008,Venegeroles2009,Avetisov2010} yield a variety of values for $\gamma_0$ ranging from 1 to 3. Some of the numerical studies have obtained different values even for the very same system. A possible explanation to this odd situation is that the long time asymptotic behavior of the Poincar\'{e} recurrence distribution cannot be reached within the existing computational power. The numerical calculation of $\gamma_0$ in the long time asymptomatic limit is exceedingly difficult. Trajectories of a particle moving in the intricate hierarchical structure of the phase space are very sensitive to numerical noise. Such a noise may transfer the particle from regions of chaotic motion to regular ones and vice versa, and can also help in crossing cantori which serve as leaky barriers within the phase space. These uncontrolled artifacts, apparently, hinder the identification of the true asymptotic behavior of $F(t)$. This explanation, however, relies on the assumption that even in the absence of numerical noise the convergence of the Poincar\'{e} recurrence distribution to its asymptotic behavior is very slow. The aim of the present work is to study this aspect of the relaxation problem. We show that $F(t)$ exhibits time dependent fluctuations which do not decay fast enough in time. Consequently $F(t)$ at different time intervals seems to exhibit different relaxation behavior which may be interpreted as different relaxation exponents. This behavior is demonstrated in Fig.~1 where the numerical calculation of $F(t)$ for the standard map is depicted on a log-log scale. It shows that different time intervals may be associated with different decay exponents. Thus convergence to the asymptotic behavior is extremely slow, and not monotonous, but rather oscillatory in $\log t$.

\begin{figure}[tbh] \centerline{\resizebox{0.47 \textwidth}{!} {\includegraphics{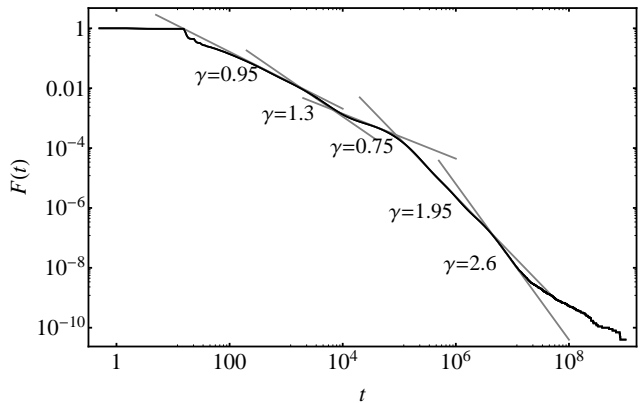}}} \caption[]{A log-log plot of the survival probability of the standard map showing local fluctuations in its decay power $\gamma$.} \label{main1} \end{figure}

These variations in the relaxation dynamics emerge from random-like local variations in the phase space of the chaotic component of mixed systems. The trajectories that contribute to $F(t)$ as time progresses are trajectories which become closer to the boundaries of the regions with regular motion. The closer the trajectory approaches to such a boundary, the longer it sticks to it. This behavior implies that self-averaging  of the relaxation dynamics is not very effective and therefore the convergence to the asymptotic value of the decay exponent is rather slow as demonstrated in Fig.~1.

Obviously, a direct study of the fluctuations of $F(t)$ around its asymptotic behavior (\ref{power-law}),  whether analytical or numerical, suffers from the same difficulties of the calculation of the asymptotic decay exponent itself. To circumvent this difficulty it is convenient to introduce an ensemble of mixed chaotic systems and to use ensemble averaging in order to extract the properties which characterize the relaxation dynamics. This is similar in spirit to disorder averaging\cite{Abrikosov} which allows one to characterize the dynamics of a particle in disordered system, such as the correlation function of its position at different times. However, unlike disordered systems where the meaning of their ensemble is well understood, it is unclear what might be the invariant measure of the ensemble of mixed chaotic systems. Nonetheless, the role of the ensemble averaging which we shall employ here is merely a regularization procedure which allows one to extract the intrinsic properties of the pure system. It is similar to the introduction of an infinitesimal noise into the dynamics of a hard chaotic system in order to extract the Pollicott-Ruelle resonances which characterize the relaxation of the pure system\cite{Gaspard2002}. Thus here we shall use disorder diagrammatics in order to perform averages and identify the correlation function of the fluctuations in the return probability. Then we shall use this result to calculate the typical value of the local fluctuations in the decay exponent $\delta\gamma = \gamma-\gamma_0$, and show that
\begin{equation}
\langle \delta \gamma^2\rangle \sim  t^{-2 \epsilon} f(\log t)  \label{deltaGammamain}
\end{equation}
where $\epsilon \ll 1$, and $f(\log t)$ is an oscillatory function.

Our analytical analysis of the problem is based on the rate equation model for the dynamics of mixed chaotic systems developed by Meiss and Ott \cite{MeissOtt}. In Sec. II we present the model where following Cristadoro and Ketzmerick\cite{Cristadoro2008} we add a small random component to the transition rates of the original  Meiss-Ott model. In section III we present the solution of the pure model, and in section IV we use the results of disorder diagrammatics to calculate the correlation function of the fluctuations in the return probability. This correlation function will be used in Sec. V in order to derive formula (\ref{deltaGammamain}) for the local fluctuations in the decay exponent. In Sec. VI we shall present numerical calculations which support our analytical results, and conclude in Sec. VII.  The technical details of our calculations can be found in the Appendices.

\begin{figure}[tbh] \centerline{\resizebox{0.47 \textwidth}{!} {\includegraphics{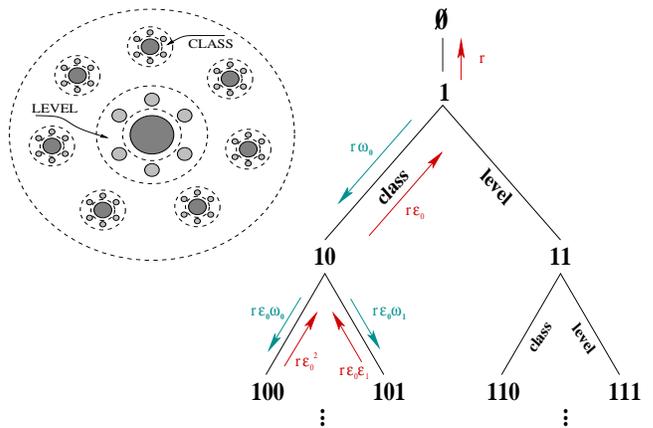}}} \caption[]{The hierarchical structure of the phase space of mixed chaotic systems (left panel), and the Meiss-Ott tree model (right panel). } \label{Heirarchical} \end{figure}

\section{The random tree model}
In this section we present our model for the ensemble of mixed chaotic systems. It is based on the assumption that the statistical characteristics of the dynamics of mixed chaotic systems can be described by rate equations. First, let us recall the rate equations approach for mixed chaotic systems introduced by Meiss and Ott\cite{MeissOtt}. Their model assumes that the dynamics within the chaotic component of the phase space can be approximated by the Master equation for the probabilities of finding the particle in states which respect the self-similar structure of the phase space, see illustration on the left panel of Fig.~2. Each one of the Meiss-Ott states is associated with a definite region of the phase space whose boundaries (represented by the dashed lines in the left panel of Fig.~2) are determined by lowest flux cantori encircling the relevant set of regular islands.

The topology of the phase space division is that of a Cayley tree, as shown in the right hand side of Fig.~2. Namely, a particle in a given state might move either to a single lower state of the hierarchy or to one of two possible higher states. A "level" transition is associated with transition to a state which is closer to the boundary circle of the island chain it is presently revolving around, while a "class" transition corresponds to the case where the particle moves into a higher state associated with one of its present satellite island chains, see Fig.~2.

The binary structure of Meiss-Ott tree allows one to designate the states of the system by binary numbers: A state reached by level transition is denoted by the number of the previous state to which a figure "1" is added, while a that reached by class transition is obtained by adding the figure "0", as demonstrated in the right hand side of Fig.~2. Let $n$ represent some arbitrary state on the tree, and denote by $Dn$ the nearest state at the upper part of the tree from which the particle may arrive. We shall also denote by $n0$ and  $n1$  the two states down the tree reached by class and level transitions, respectively. With these definitions, the Master equation takes the form
\begin{eqnarray} \frac{dP_{n}}{dt}= -\left( W_{n\to Dn}+ W_{n\to n0} +W_{n\to n1} \right) P_n  \nonumber \\ + W_{n0 \to n} P_{n0}+ W_{n1 \to n} P_{n1}+ W_{Dn \to n} P_{Dn}, \label{eq:Master}
\end{eqnarray}
where $P_n$ is the probability to find the particle in the state $n$, while $W_{n \to m}$ is the transition rate from the state  $n$ to $m$. Meiss and Ott assumed that these transition rates satisfy a simple scaling behavior:
\begin{eqnarray} \frac{W_{n \to n0}}{W_{Dn \to n}} = \varepsilon_0~ ~~~\frac{W_{n \to n1}}{W _{Dn \to n}} = \varepsilon_1 \label{Ratio1}
\end{eqnarray}
and
\begin{eqnarray} \frac{W_{n0 \to n}}{W_{n \to n0}} =  \frac{\omega_0 }{\varepsilon_0}~~~~~ \frac{W_{n1 \to n}}{W_{n \to n1}} = \frac{\omega_1}{\varepsilon_1} \label{Ratio2}
\end{eqnarray}
where $\varepsilon_0$, $\varepsilon_1$, $\omega_0$, and $\omega_1$, are constants which have been estimated to be\cite{MeissOtt}:
\begin{eqnarray}
\varepsilon_0\simeq 0.143,~~~~ \varepsilon_1 \simeq 0.382,~ \nonumber \\ \omega_0\simeq 0.0142,~~~ \omega_1 \simeq 0.0532,   \label{parameters}
\end{eqnarray}
Within this model it is also assumed that a particle staying in the upper state of the tree leaves the tree at rate $r$ and never return back. From here on, the escape rate $r$ will be set to unity by choosing proper units of time.

It is known, however, that the rates ratios (\ref{Ratio1}) and (\ref{Ratio2}) fluctuate considerably at different positions of the tree\cite{Green86}. Therefore it is natural to generalize the Meiss-Ott model by adding a random component to the transition rates, i.e. to replace the constant transition rates by fluctuating ones:
\begin{eqnarray} W_{n \to m} \to  W_{n \to m}  (1+\xi_{nm}) \label{RandomTransitionRates}
\end{eqnarray}
where $\xi_{nm}$ are assumed to be uncorrelated random variables with zero mean,
\begin{eqnarray}
\langle \xi_{nm} \xi_{n'm'} \rangle = \sigma^2 \left( \delta_{nn'}\delta_{mm'}+ \delta_{nm'}\delta_{mn'}  \right). \label{xi-average}
\end{eqnarray}
Here $\sigma$ is a dimensionless constant which controls the amount of randomness of the ensemble.

This ensemble is based on two main assumptions: (a) The main source of fluctuations comes from the flux exchange between states, while deviations from the scaling behavior of phase space areas of the states may be neglected. (b) The fluctuations in the flux exchange through different boundaries of the states are uncorrelated. The above choice implies that statistical properties of the dynamics on small scales of the phase space is the same as at larger scales with the proper rescaling of time. We shall refer to this ensemble as the "random tree model".

In order to identify the relation between the functions $P_n(t)$ and the survival probability, $F(t)$, let us sum the Master equation (\ref{eq:Master}) over all the states of the tree $n$:
\begin{eqnarray} \frac{d}{dt} \sum_n P_n= \frac{dF(t)}{dt} = - P_1 \label{sur-ret-rel}
\end{eqnarray}
The left hand side of this equation is precisely the derivative of the survival probability, $F(t)= \sum_n P_n$, since the sum over $P_n$ is the probability to find the particle at any site on the tree. From here and  (\ref{power-law}) it follows that the long time asymptotic decay of $P_1(t)$ is
\begin{eqnarray}
P_1(t) \sim t^{-1-\gamma_0}.   \label{LTAB}
\end{eqnarray}
Thus one may study the convergence to this asymptotic limit by analyzing the behavior of the return probability
\begin{eqnarray}
P(t) =P_1(t)   ~~ \mbox{where}~~  P_n(0) =\delta_{n,1}  \label{RetProbDef}
\end{eqnarray}
This is the probability density that a particle is found at upper site of the tree, $n=1$ assuming that
 at $t=0$ the particle is placed at the same site.

 \section{Solution of the nonrandom model}
 In this section we review the solution of the Meiss-Ott model that describes the pure system,
 thereby presenting some of the ingredients which will serve us in the next section where we consider the random model.
 Let us define the probability $P_{n,m}(t)$  of a particle, initially placed at site $m$,  to be at site $n$ at time $t$.
 We define the Green function, $\tilde{G}_{n,m}(s)$, as the Laplace transform of $P_{n,m}(t)$,
\begin{eqnarray} \tilde{G}_{n,m}(s)= \int_0^\infty dt e^{-s t}P_{n,m}(t). \label{GFDef}
\end{eqnarray}
Our aim is to characterize the analytic structure of the green function associated with the return probability to the upper site of the tree, $\tilde{G}_{1,1}(s)$. In particular we shall focus our attention on the analytic structure in the vicinity of the point $s=0$ which governs the long time asymptotic behavior of the return probability (\ref{RetProbDef}). In this vicinity one expects  $\tilde{G}_{1,1}(s)$ to take the form
\begin{eqnarray}
\tilde{G}_{1,1}(s) = a(s)+ b(s) s^{\gamma}+ \cdots \label{tildeexp}
\end{eqnarray}
where $a(s)$ and $b(s)$ are some analytic functions. Since the
inverse Laplace transform of an analytic function decays, in time, faster than any power law, the long time asymptotic behavior (\ref{LTAB}) comes from the nonanalytic contribution represented by the second term of the expansion (\ref{tildeexp}). Meiss and Ott proved that $\gamma$ satisfies the dispersion equation\cite{MeissOtt}:
\begin{eqnarray}
\omega_0 \varepsilon_0^{-\gamma} + \omega_1 \varepsilon_1^{-\gamma}=1. \label{eq:gamma1}
\end{eqnarray}
This equation is obtained by substituting (\ref{tildeexp}) into the equation that the Green function $\tilde{G}_{1,1}(s)$ satisfies. It is rederived  using a diagrammatic approach in Appendix A.

\begin{figure}[tbh] \centerline{\resizebox{0.4 \textwidth}{!} {\includegraphics{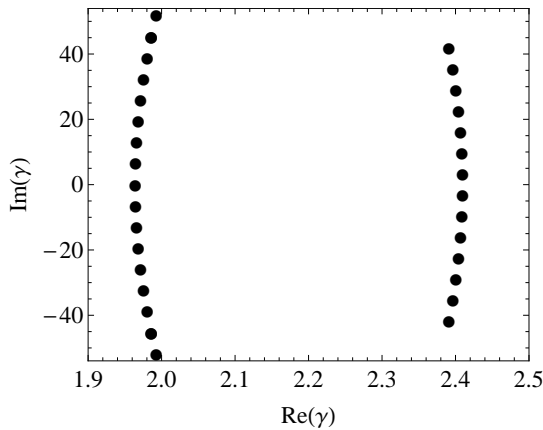}}} \caption[]{The solutions of the dispersion equation (\ref{eq:gamma1})} \label{GammaEigen}
\end{figure}

Let us examine the solutions of the dispersion equation (\ref{eq:gamma1}). Apart from a single purely real solution the other solutions of this equation appear in complex conjugate pairs. In Fig.~3 we present some of these solutions in the complex plain $(\mbox{Re} (\gamma), \mbox{Im} (\gamma))$. From here it follows that the solutions are divided into two groups. In each one of these groups, the real parts of the solutions are approximately the same. The difference between the solutions is mainely due to their imaginary parts. The solution with smallest real part is
\begin{eqnarray}
\gamma_0 = 1.96424 \label{gamma-infinity}
\end{eqnarray}
This solution governs the asymptotic long time behavior. The solutions with the next smallest real part are  $\gamma_1 =1.96469\pm i 6.47555$, while the next solutions are $\gamma_2=1.96605\pm i 12.9511\imath$. Thus the general expansion of the Green function near $s=0$ is
\begin{eqnarray}
\tilde{G}_{1,1}(s) = a(s)+ \sum_\nu b_\nu(s) s^{\gamma_\nu}+ \cdots \label{g-expansion1}
\end{eqnarray}
The form of decay associated with complex solutions of the form $s^{\gamma_R \pm i \gamma_I}$ is $t^{-\gamma_R} \cos(\gamma_I \log t + \phi)$ where $\phi$ is some constant. Thus neglecting the difference in the real parts of the solutions we see that the long asymptotic behavior of the return probability of the pure system is
\begin{eqnarray}
\tilde{P}(t) = P_1(t) \sim t^{-\gamma_0-1} h_1( \log t) \label{Pt}
\end{eqnarray}
where $h_1(x)$ is non-periodic oscillatory function whose behavior depends on the initial conditions of the system. $h_1(x)$  may be approximated by the leading solutions of the dispersion equation (\ref{eq:gamma1}), i.e.
\begin{eqnarray}
h_1(x) \approx  \alpha_0+ \alpha_1 \cos(\gamma' x + \phi) \label{h1}
\end{eqnarray}
where $\alpha_0$, $\alpha_1$ and $\phi$ are positive constants, and $\gamma' =6.47555$ is the imaginary part of the next to  leading solution of the dispersion equation (\ref{eq:gamma1}).
\\ \ \\
\section{Solution of the random tree model}
While ensemble averaging is not required for extracting the long time asymptotic decay of the return probability, the sample to sample fluctuations are defined only with regard to ensemble. In this section our goal is to calculate the correlation function of the sample to sample fluctuations:
\begin{eqnarray}
{\cal C}(t, t')= \langle \delta P(t) \delta P(t') \rangle \label{correlation-fluc}
\end{eqnarray}
where
\begin{eqnarray}
\delta P(t)= P(t)- \langle P(t)\rangle
\end{eqnarray}
is the deviation of the return probability of the system with the random component, $P(t)$, from the average over the ensemble $\langle P(t)\rangle$.

Since we employ the ensemble averaging procedure only as a tool for extracting the intrinsic properties of the system, we may assume that fluctuations in the transition rates are small, $\sigma \ll 1$, and exploit $\sigma$ as the small parameter of the perturbation theory. The details of this perturbation theory are presented in Appendix B.

The correlation function (\ref{correlation-fluc}) may be written as a double inverse Laplace transform \begin{eqnarray}
{\cal C}(t,t') = \int_{c-i\infty}^{c+ i \infty}\!\frac{ds'}{2\pi i} \int_{c-i\infty}^{c+ i \infty}\!\frac{ds}{2\pi i} e^{s t+s't'} Q(s,s')\label{eq:convolution}
\end{eqnarray}
where the constant $c$ in the integration contour is set to ensure the convergence of the integral, and $Q(s',s)$ is the disconnected part of the correlation of the Green function
\begin{eqnarray}
Q(s',s)= \langle  G_{1,1}(s')  G_{1,1}(s)\rangle- \langle  G_{1,1}(s') \rangle \langle  G_{1,1}(s)\rangle,  \label{TGF-correlator}
\end{eqnarray}
where  $G_{1,1}(s')$ denotes the Green function of the random tree model. In appendix B it is shown that, to the lowest order in  $\sigma$,  $Q(s's)$ stratifies the equation:
\begin{widetext}
\begin{eqnarray}
 Q(s',s)= \Lambda (s',s) +  \tilde{G}_{1,1}^2(s') \tilde{G}_{1,1}^2(s)  \left[\omega_0^2 Q\left(\frac{s'}{\varepsilon_0},\frac{s}{\varepsilon_0}\right)+\omega_1^2 Q\left(\frac{s'}{\varepsilon_1},\frac{s}{\varepsilon_1}\right) \right]      \label{Q-eq}
 \end{eqnarray}
 \end{widetext}
 where
 \begin{eqnarray}
 \Lambda (s',s)= \sigma^2 \tilde{G}_{1,1}^2(s') \tilde{G}_{1,1}^2(s) \left[ 1+ \sum_{j=0,1} \omega_j^2 \alpha_j(s',s) \right] \label{LambdaFor}
 \end{eqnarray}
 and $\alpha_j(s',s)$ is a function expressed in terms of the Green function of the pure system, $\tilde{G}_{1,1}(s)$. Its explicit form can be found in Appendix B.

 From the structure of Eq. (\ref{Q-eq}) it follows that its solution contains two contributions: The homogeneous solution and the inhomogeneous one. Consider first the inhomogeneous solution which we denoted by $Q_{in}(s's)$. Focusing our attention on the terms which are relevant for the long time asymptotic behavior we expand this solution in the form
 \begin{widetext}
 \begin{eqnarray}
 Q_{in}(s',s)= q^{(0)}(s',s) + \sum_\nu \left(q_\nu^{(1)}(s',s) s^{\gamma_\nu-1} +  q_\nu^{(1)}(s,s') s'^{\gamma_\nu-1} \right)+ \sum_{\nu,\nu'} q^{(2)}_{\nu,\nu'}(s',s)s'^{\gamma_{\nu'}-1} s^{\gamma_\nu-1} + \cdots \label{Q-expansion}
 \end{eqnarray}
 \end{widetext}
 where $q^{(j)}(s',s)$ represent functions that are analytic in $s$ and $s'$, and $\gamma_\nu$ are the solutions of Eq. (\ref{eq:gamma1}). Substituting this form in (\ref{Q-eq}) it is evident that with a proper choice of the functions $q^{(j)}(s',s)$ one can satisfy the equation term by term. The first two terms of the expansion (\ref{Q-expansion}) are analytic either in $s$ or in $s'$ and therefore do not contribute to the long time asymptotic behavior of the fluctuations. The third term, near $s=s'=0$, may be written in the form of a product: $\chi \sigma^2 (\sum_\nu b_\nu(0) s^{\gamma_\nu-1})(\sum_\nu b_\nu(0) s'^{\gamma_\nu-1})$ where $b_\nu(0)$ are the expansion coefficients of the Green function (\ref{g-expansion1}), and $\chi$ is a constant which may be expressed in terms of $a(0)$, and the parameters (\ref{parameters}). This implies that the contribution to the correlation function of the fluctuations in the long time asymptotic limit is proportional to the product of the return probabilities:
 \begin{eqnarray}
 {\cal C}_{in}(t,t') = \chi \sigma^2   \tilde{P}(t) \tilde{P}(t') \label{deltaP2in}
 \end{eqnarray}
 where $\tilde{P}(t)$ is the return probability of the system without the random component, see Appendix A.

 Consider now the homogeneous solutions of Eq. (\ref{Q-eq}), which we shall denote by $Q_h(s's)$. The total solution of the equation is $Q(s',s)= Q_{in}(s's)+Q_h(s's)$ and its inverse Laplace transform gives the correlation function ${\cal C}(t,t')$-see Eq. (\ref{eq:convolution}). The homogeneous solution is needed in order to satisfy the condition
 \begin{eqnarray}
 {\cal C}(0,t')={\cal C}(t,0)=0 \label{cond1}
 \end{eqnarray}
 for any $t$ and $t'$. This is because at the initial time of evolution $\delta P(0)=0$ as the system is prepared in such a way that the particle is with probability one at the upper state of the tree,  $"1"$. To obtain the long time asymptotic behavior of the homogeneous solution, we expand $Q_h(s',s)$ in the form
 \begin{widetext}
 \begin{eqnarray}
 Q_h(s',s)=  v^{(0)} (s',s)+\sum_\nu \left(v^{(1)}_\nu (s',s)s^\mu_\nu+v^{(1)}_\nu (s,s')s'^\mu_\nu\right) + \sum_{\nu,\nu'}v^{(2)}_{\nu,\nu'}(s',s) s^{\mu_\nu} s'^{\mu_{\nu'}}+ \cdots \label{Q-expansion2}
  \end{eqnarray}
  \end{widetext}
 where $v^{(j)}(s',s)$ are analytic functions at $s=s'=0$ and $\mu_\nu$ are unknown exponents. Substituting this expansion in the homogeneous equation obtained from (\ref{Q-eq}), and solving the resulting equation term by term, one obtains that the contribution associated with the slowest decay exponents (which comes from the third term in the expansion (\ref{Q-expansion2})) should satisfy the dispersion equation:
 \begin{eqnarray}
 \left(\omega_0 \varepsilon_0^{-\bar{\mu}}\right)^2 +\left( \omega_1 \varepsilon_1^{-\bar{\mu}}\right)^2 = 1 \label{mu3}
 \end{eqnarray}
 where
 \begin{eqnarray}
  \bar{\mu}= \frac{\mu_\nu+ \mu_{\nu'}}{2} \label{muav}
 \end{eqnarray}
 This equation, similar to Eq. (\ref{eq:gamma1}), has many solutions for $\bar{\mu}$ which form two branches similar to those which appear in Fig.~3. The solution with the smallest positive real part satisfies the relation $\bar{\mu}_0 > \gamma_0$, and for the parameters (\ref{parameters}) its value is $\bar{\mu}_0=2.13852$. Other exponents belonging to the same branch have the approximate form $\bar{\mu}_0 \pm  i \mu'_j$ for example $\mu'_1 = 3.23377$ and  $\mu'_2= 6.46753$. Since the dispersion equation (\ref{mu3}) constrains only the sum of exponents (\ref{muav}) there is apparently a continuous set of solutions associated with different values of the difference between the exponents,  $\mu_\nu- \mu_{\nu'}$, the weight of each one of these solutions is determined by $v^{(2)}_{\nu,\nu'}(s',s)$. This approach does not allow us to determine these weights since our solution is valid in the long time asymptotic limit while the condition (\ref{cond1}) corresponds to short time dynamics.  Nevertheless, we may express the homogeneous contribution to the correlation function, in the long time asymptotic limit, as a sum
 \begin{eqnarray}
 {\cal C}_{h}(t,t') = \mbox{Re} \sum_{\nu,\eta} c_{\nu, \eta}  \left( t^{-\eta} t'^{2\bar{\mu}_\nu- \eta} +t'^{-\eta} t^{2\bar{\mu}_\nu- \eta} \right)
 \end{eqnarray}
 where $c_{\nu,\eta}$ are some unknown constants and $\bar{\mu}_\nu$ are the solution of the dispersion equation (\ref{mu3}). Here we assume that the sum is only over the solutions with positive imaginary part. In particular for $t'=t$ one obtains
 \begin{eqnarray}
 {\cal C}_{h}(t,t) =  \langle \delta P(t)^2 \rangle_h \simeq \sigma^2 t^{-2\bar{\mu}_0} h_2( \log t) \label{delta2Pt}
 \end{eqnarray}
 where $h_2(\log t)$ is an oscillatory function  which similarly to $h_1(t)$ (see Eq. (\ref{h1}))may be approximated in the long time asymptotic limit as
 \begin{eqnarray}
 h_2(x) \approx  d_0+ d_1 \cos(2\mu'_1 x + \phi_1)+d_3 \cos(2\mu'_2 x + \phi)+ \cdots \label{h2}
 \end{eqnarray}
 where $d_j$ and $\phi_j$ are some constants, and $\mu'_j$ represent the imaginary part of the solutions of the dispersion equation  (\ref{mu3}).

 Now from  (\ref{deltaP2in}), (\ref{delta2Pt}) and the results presented in Sec. V we get that the variance of the normalized fluctuations $\delta p(t)= \delta P(t)/ \langle P(t) \rangle$, in the long time asymptotic limit, and to the leading order in $\sigma$ (where $\langle P(t) \rangle$ should be replaced by $\tilde{P}$)takes the form
 \begin{eqnarray}
 \left\langle \delta p^2\left( t \right)\right\rangle = \chi \sigma^2 \left( 1+ \beta_1(t) \right)  \label{fluct}
 \end{eqnarray}
 where
 \begin{eqnarray}
 \beta_1(t) = \frac{t^{-2\epsilon}}{\chi}  \frac{h_2(\log t)}{h_1^2(\log t)}  \label{beta1}
 \end{eqnarray}
 and
 \begin{eqnarray}
 \epsilon= \bar{\mu}_0-\gamma_0  \simeq 0.174 \label{main-epsilon}
 \end{eqnarray}
 Thus the normalized fluctuations in the return probability decay to a constant value very slowly and in an oscillatory manner as function of $\log t$. Finally, using the above results for the homogeneous and inhomogeneous contribution to the correlator ${\cal C}(t+ \Delta t/2), t-\Delta t/2)$, and expanding it to second order in $\Delta t/t$, we obtain:
 \begin{widetext}
 \begin{eqnarray}
 {\cal C}\left(t+ \frac{\Delta t}{2}, t-\frac{\Delta t}{2}\right) \simeq \sigma^2 \left( \chi  t^{-2\gamma_0}\left( h_1^2(\log t) + \tilde{h}_1(\log t) \left(\frac{\Delta t}{2t} \right)^2 \right)+ t^{-2\mu_0}\left( h_2(\log t) + \tilde{h}_2(\log t) \left(\frac{\Delta t}{2t} \right)^2 \right) \right) \end{eqnarray}
 \end{widetext}
 where
 \begin{eqnarray}
 \tilde{h}_1(x)=  \gamma_0  h_1^2(x)\!-\! h_1'^2(x)\!+\! h_1(x)( h''_1(x)\!-\! h'(x)) \label{htilde}
 \end{eqnarray}
 and is $\tilde{h}_2(\log t)$ is an oscillatory function with expansion similar to (\ref{h2}) but with different coefficients.  From here one obtains the correlator of the normalized fluctuations
 \begin{equation}
 \frac{\left\langle \delta p\left(t- \frac{\Delta t}{2}\right) \delta p \left(t+ \frac{\Delta t}{2}\right) \right\rangle}{\left\langle \delta p^2\left( t \right)\right\rangle}  \simeq   1-\frac{1}{2} \beta_2(t) \left(\frac{\Delta t}{t}\right)^2  \label{rmain}
 \end{equation}
 where
 \begin{eqnarray} \beta_2 (t) =  t^{- 2\epsilon}  \frac{ \gamma_0 h_2 (\log t)-\tilde{h}_2(\log t) }{2 \chi h_1^2(\log t)} \label{beta2}
 \end{eqnarray}
 To obtain this result we take into account that $h_1(x)$ is approximately constant and thus $\tilde{h}_1(x)$ may be approximated by the first term on the right hand side of (\ref{htilde}).

 \section{Sample to sample fluctuations of the decay exponent}

 In this section we calculate the typical value of the sample to sample fluctuation in the local behavior the decay exponent $\gamma$. For this purpose let us present the return probability in the form
 \begin{equation}
 P(t) = \langle P (t) \rangle (1 + \delta p(t)),
 \end{equation}
 where $\langle P (t) \rangle$ is the long time asymptotic limit of the average of the return probability over the ensemble which to the leading order in the disorder may be taken as the return probability of the pure system,  (\ref{Pt}). The function $\delta p(t) = P(t)/\langle P (t) \rangle  -1$ is the normalized sample to sample fluctuation which is characterized by the correlation function (\ref{rmain}). Assume one wishes to extract the decay exponent $\gamma$, of some particular system, from the measurement of $P(t)$ at two time points, say $t\pm \Delta t/2$ where the time difference $\Delta t$ between the measuring points is assumed to be smaller than $t$, $\Delta t \ll t$ (as is the general situation when one tries to extract the limiting value of the decay exponent from the tail of $P(t)$) . Assuming that within this range the local behavior is  $P(t)\propto t^{-1-\gamma}$, one obtains that
 \begin{eqnarray}
 \gamma =1+\gamma_0+ \frac{t}{\Delta t} \ln \left(\frac{1+\delta p(t- \Delta t/2)}{1+ \delta p(t+\Delta t/2)}\right),
 \end{eqnarray}
 Thus $\delta \gamma= \gamma -\gamma_0$ can be associated with the normalized fluctuations of the return probability, $\delta p(t)$. Expanding the above logarithm to the leading order in $ \delta p(t)$, one may express the mean square of the fluctuations in the decay exponent in the form
 \begin{equation}
 \left\langle \delta \gamma^2\right\rangle \!\simeq\! \left(\frac{ t}{\Delta t}\right)^2 \left\langle \!\left( \delta p\left(t\!+\! \frac{\Delta t}{2} \right)\!-\! \delta p\left( t \!-\!\frac{\Delta t}{2}\right) \right)^2 \! \right\rangle,
 \end{equation}
 Using (\ref{rmain}) and (\ref{fluct}) we obtain
 \begin{equation}
 \langle \delta \gamma^2\rangle \simeq  \chi \sigma^2 \left(\beta_2 (t)+ t^2\beta_1''(t) \right)  \label{avdseltasquare}
 \end{equation}
 where  $\beta_1(t)$ and $\beta_2(t)$ are given by (\ref{beta1}) and (\ref{beta2}) respectively.   Both functions are oscillatory in $\log t$ and decay as $t^{-2\epsilon}$. Thus the typical fluctuation in the value of decay exponent decays as $\delta\gamma \sim t^{-\epsilon}$, where $\epsilon \approx 0.174$. From here one concludes that the fluctuations decay very slowly in time and in an oscillatory manner.

 \section{Numerical study}
 In this section we present our numerical study which has the following goals: First to verify our analytical solution. Second to show that the results that we have obtain in the limit $\sigma \ll 1$ are insensitive to the magnitude of the random component of the transition rates. Third, to compare our results with those obtained from the exact dynamics of  an ensemble of symplectic maps.

 Our numerical study of the random tree model is performed directly for the survival probability, $F(t)$, rather than the return probability $P(t)$. The computations are performed using 1000 realizations of the ensemble of random tree model with 9 generations. The random variables $\xi_{n,m}$, are chosen form a uniform distribution.

 The relation (\ref{sur-ret-rel}) implies that $F(t)$ behaves in a similar way to  $P(t)$. In particular one expects that
 \begin{eqnarray} F(t)= t^{-\gamma_0} h(\log t) \label{eq:F}
 \end{eqnarray}
 where similar to (\ref{Pt}), the function $h(\log t)$ is an oscillatory function whose expansion is the same as that of $h_1(\log t)$  but with different coefficients. Yet, formula (\ref{h1}) describes the behavior in the very long time asymptotic behavior, while within the time limits were our calculation is reliable one has to take into account also terms associated with the second branch of solutions of the dispersion equations (\ref{eq:gamma1}) that are associated with faster decaying rates (see right branch of solutions in Fig.~3).

 The numerical calculation of $h(\log t)$ for the pure tree model is presented by the black dots in Fig.~4.
 This solution is obtained by calculating the normal modes of the tree model and expanding the solution in these modes. The solid line in this figure represents the analytical solution described by the function
 \begin{eqnarray}
 h(x) \simeq   \alpha_0 &+& \frac{\alpha_1}{t}+ \alpha_2 t^{\gamma_0-\gamma'_1} \cos(\gamma''_1 x + \phi_1) \nonumber \\ &+&  \alpha_3 t^{\gamma_0-\gamma'_2}\cos(\gamma''_2 x + \phi_2) \label{h}
 \end{eqnarray}
 where $\gamma'_1\pm i \gamma''_1= 1.9646\pm i 6.4755$ and $\gamma'_2\pm i \gamma''_2= 2.4097\pm i 3.223$ are the slowest oscillatory solutions of the dispersion equation (\ref{eq:gamma1}) while the term $\alpha_1/t$ comes from the linear in $s$ term of the expansion of the function $a(s)$ of the Green function (\ref{tildeexp}). The parameters $\alpha_j$ and $\phi_j$ are fitting parameters. The analytical solution gives an excellent fit to the numerical data except at very large times where numerical errors are large. Notice that the amplitude of the oscillatory component of $h(t)$ is very small, thus to good approximation it may be considered to be constant.

 \begin{figure}[tbh]
 \centerline{\resizebox{0.49 \textwidth}{!} {\includegraphics{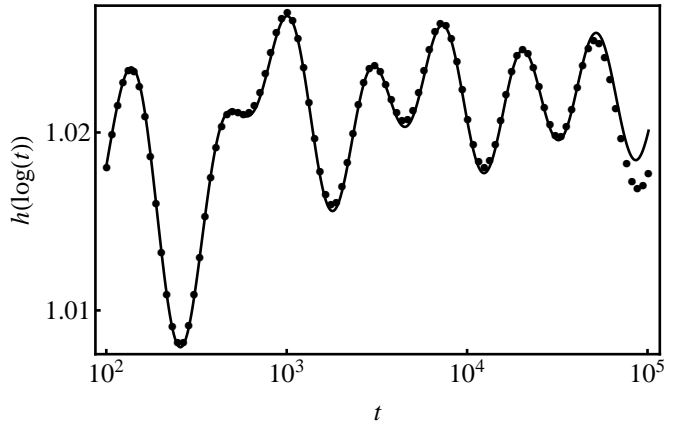}}} \caption[]{The numerical (black dots) and analytical (solid line) results for the function $h(t)$ which dictates the oscillatory nature of the survival probability  (\ref{eq:F}).}
 \end{figure}

 Let us consider the behavior of  $\langle \delta\gamma^2 \rangle$. From formula (\ref{avdseltasquare}), expressions (\ref{beta1}) and (\ref{beta2}), and taking into account that to a good approximation $h(t)$ may be replaced by constant, it follows that $t^{2\epsilon} \langle \delta \gamma^2 \rangle$ may be expanded in terms of the solutions $\mu' \pm i  \mu"$, of the dispersion equation (\ref{mu3}), namely \begin{eqnarray}
 t^{2\epsilon} \langle \delta \gamma^2 \rangle = t^{ 2\bar{\mu}_0} \sum_j v_j t^{- \mu'_j} \cos (\mu"_j \log t - \phi_j) \label{analytical}
 \end{eqnarray}
 The black dots in Fig.~5 represent the numerical calculation of this function for the case were $\sigma=0.04$, while the solid line is the analytical expression (\ref{analytical}) in which we took the four slowest oscillatory solutions of (\ref{mu3}) and use  $v_j$ and $\phi_j$ as fitting parameters. The very good agrement between the numerical and analytical results shown in Fig. 5 proves that the typical amplitude of the fluctuation, $\delta \gamma$ decays as $t^{-\epsilon}$ where $\epsilon$ is given by (\ref{main-epsilon}). It also shows that the oscillatory behavior of $\langle \delta \gamma^2 \rangle$ is dictated by the oscillatory solutions of (\ref{mu3}).

 \begin{figure}[tbh]
 \centerline{\resizebox{0.47 \textwidth}{!} {\includegraphics{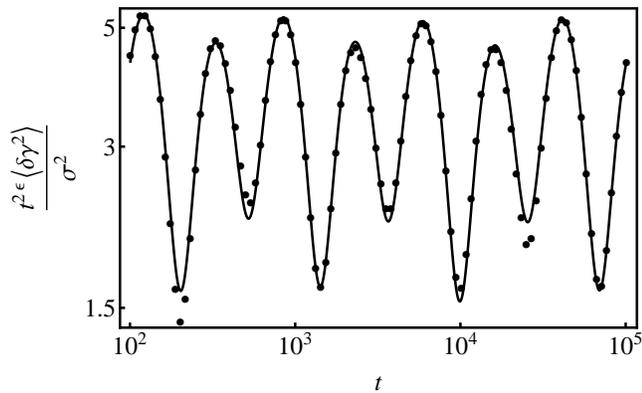}}} \caption[]{The numerical (black dots) and analyticl (solid line) results for the oscillatory factor of the sample to sample fluctuations of the decay exponent (\ref{analytical}).}
 \end{figure}

 Our analytical results have been derived for weak disorder. In this limit the strength of the random component in the transition rates appears only through the prefactor which controls the magnitude of the fluctuation (see Eq. (\ref{avdseltasquare})), while the functional dependence on time is independent of the $\sigma$ and depends only on the intrinsic properties of the pure Meiss-Ott model. Our numerical comparison, shown in Fig.~5, has been also performed for weak disorder $\sigma = 0.04$. It is, however, worthwhile to clarify to what extent these results depend on the value of $\sigma$.  . To this end we compute  $\langle \delta \gamma^2 \rangle$ for various values of $\sigma$, between $0.04$ and $0.2$ (Notice that the higher value is not far from the upper limit of the widest possible uniform distribution for which $\sigma=1/\sqrt{12}\simeq 0.29$).

 The functions $t^{2\epsilon} \langle \delta \gamma^2 \rangle/\sigma^2$ are presented in Fig.~6. The fact that they almost collapse on the same graph, implies that our results are almost independent of the strength of the random component of the transition rates.

\begin{figure}[tbh] \centerline{\resizebox{0.47 \textwidth}{!} {\includegraphics{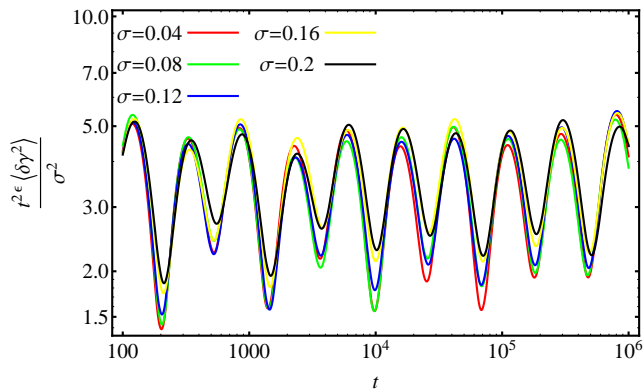}}} \caption[]{(Color online) The functional behavior of the sample to sample fluctuations of the random tree model for various values of the disorder strength.}
\end{figure}

Finally it is instructive to compare our results with those of the true dynamics of mixed chaotic systems. For this purpose we study an ensemble of chaotic maps obtained by adding a small random component to the standard map:
\begin{eqnarray} I_{n+1} &=& I_n + K \sin(\theta_n)+ {\cal R}(\theta_n) \nonumber \\ \theta_{n+1} &=& \theta_{n}+I_{n+1} \;\mbox{mod} (2\pi) \label{standard-map}
\end{eqnarray}
Here $K$ is the kicking strength, and ${\cal R}(\theta)$ is a Gaussian random function  with zero mean and periodic correlation
\begin{eqnarray} \langle {\cal R}(\theta){\cal R}(\theta') \rangle =\tilde{\sigma}^2 \sum_m e^ {- \frac{(\theta- \theta' -2 \pi m)^2 }{l^2}} \label{random}
\end{eqnarray}
The constant $\tilde{\sigma}$ controls the strength of the random contribution, while $l$ is the correlation length of ${\cal R}(\theta)$. Thus each realization of ${\cal R}(\theta)$ corresponds to a slightly different symplectic map, which may be viewed as a different member of the random tree model. Thus averaging over this ensemble is expected to have similar effects to averaging over the transition rates of the random tree model. For the numerical computation we choose $\tilde{\sigma}= 0.21\cdot10^{-5}$ which is much smaller than the largest value of $K$, for which the unstable fixed point at $I=0$ and $\theta=\pi$ remains unstable. The initial points of the various trajectories of the particles, are chosen to be in the vicinity of this fixed point (within a square of size $10^{-6} \times 10^{-6}$), and the particle is assumed to leave the system when it crosses the line $I=0$. We choose kicking strength  $K=0.971635406$,  correlation length $l=0.2$, and the calculation is performed using a quadruple precision. The map was iterated up to $5 \cdot 10^5$ iterations, which is the range within which the results agree with those obtained by double precision. For each realization of the random function ${\cal R}(\theta)$, the survival probability was calculated using  $2\cdot 10^9$ trajectories. The correlations of the fluctuation have been obtained by averaging over 260 different realizations of the random component.

In Fig.~7 we present the results for the fluctuations in the decay exponent, $\langle \delta \gamma^2 \rangle$, as function of $\log t$. This graph shows an oscillatory behavior of the fluctuations but with no apparent decay, namely $\epsilon \approx 0$.

\begin{figure}
 \includegraphics[width=8.5cm]{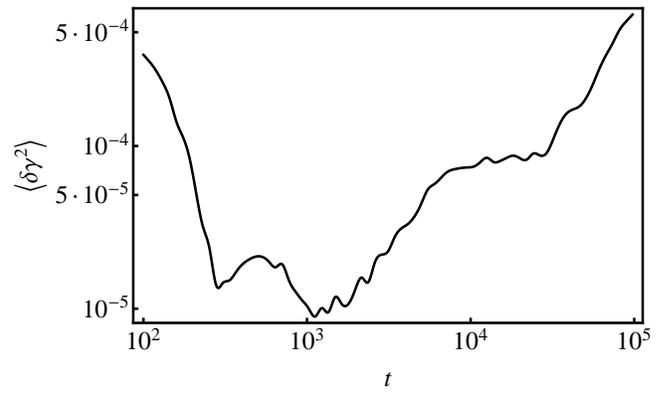} \caption{(Color online) The relative fluctuations in the distribution of Poincar\'{e} recurrences of the random standard map model (\ref{standard-map}) versus time.} \label{relativeFluctuations}
 \end{figure}

 \section{Conclusions}
 In this work we have extended the rate equation approach of Meiss-and Ott in order to calculate the functional behavior of the sample to sample fluctuations in the local decay exponent $\gamma$ of the survival probability. Our calculations were conducted within the leading order perturbation theory in the strength of the random component. In this limit, it has been shown that apart from the amplitude of the fluctuations in $\gamma$, their functional dependence on time is dictated by the intrinsic parameters of the Meiss-Ott model. Moreover, our numerical study shows that this conclusion is valid also at strong disorder. Our approach cannot give the typical value of the random component, $\sigma$. However, numerical evidence from the study of mixed chaotic systems suggests that it is of order unity. If we also assume that the typical magnitude of the fluctuations in $\gamma$ is characterized by the decay exponent $\epsilon =0.174$, then even after a million iterations the typical fluctuation in $\gamma$ is of order of one tenth.

Moreover, our numerical study of the ensemble of symplectic maps (\ref{standard-map}), does not show any sign of decay of the fluctuations. Namely, within the limits of reliable numerical calculation it is difficult to obtain the asymptotic value of the decay exponent, $\gamma_0$ (as also evident from Fig. 1). One may speculate that this is because the assumption that the random component in the transition rates is uncorrelated is not correct, and that its effect cannot be considered to be perturbative. For instance it is plausible that strong and correlated random components may confine the dynamics, in the long time limit, to a very small number of branches of the tree model. This implies that self averaging is not effective and therefore fluctuations in the normalized return probability, $\delta p(t)$ do not decay in time, or in other words  $\epsilon \simeq 0$. This result is also obtained for a very asymmetric random tree model where class transitions are much smaller than level transition $\varepsilon_0/\varepsilon_1, \omega_0/\omega_1 \to 0$. In this very asymmetric model, $\epsilon  \to 0$ and therefore $\delta \gamma$ does not decay in time.

Let us, finally, remark that the tree model of Meiss and Ott represents an uncontrolled phenomenological approximation of the exact dynamics which is based on the assumption that within the phase space region associated with a given state, the relaxation is much faster compared to the transition time to other states. The status of validity of this assumption, however, is unclear\cite{Rom-Kedar90}. Nevertheless, our analysis corresponds to a worst case scenario were self averaging is effective, and even in this case we obtain that the fluctuations in return probability decay extremely slowly in time. Thus the convergence of the survival probability, $F(t)$, to its asymptotic value is very slow. This explains the wide range of results for the decay exponent $\gamma$ obtained by numerical studies [5-18] in the last three decades.

\acknowledgments
We thank Avraham Klein and  Shmuel Fishman for valuable comments on this work. This research has been supported in part by the Israel Science Foundation  (ISF) under Grant 9/09 and by the United States-Israel Binational Science Foundation (BSF) grant No. 2008278.

 \section{Appendix A: Derivation of Eq. (\ref{eq:gamma1})}

 In this appendix we derive the dispersion
 equation (\ref{eq:gamma1}) which determines $\gamma_0$ as well as the other exponents of the relaxation modes shown in Fig.~3. For this purpose we first derive an exact equation of the Green function $\tilde{G}_{1,1}(s)$ and then show that a solution of the form (\ref{tildeexp}) leads to the dispersion equation.

\begin{figure}[tbh]
\centerline{\resizebox{0.45 \textwidth}{!} {\includegraphics{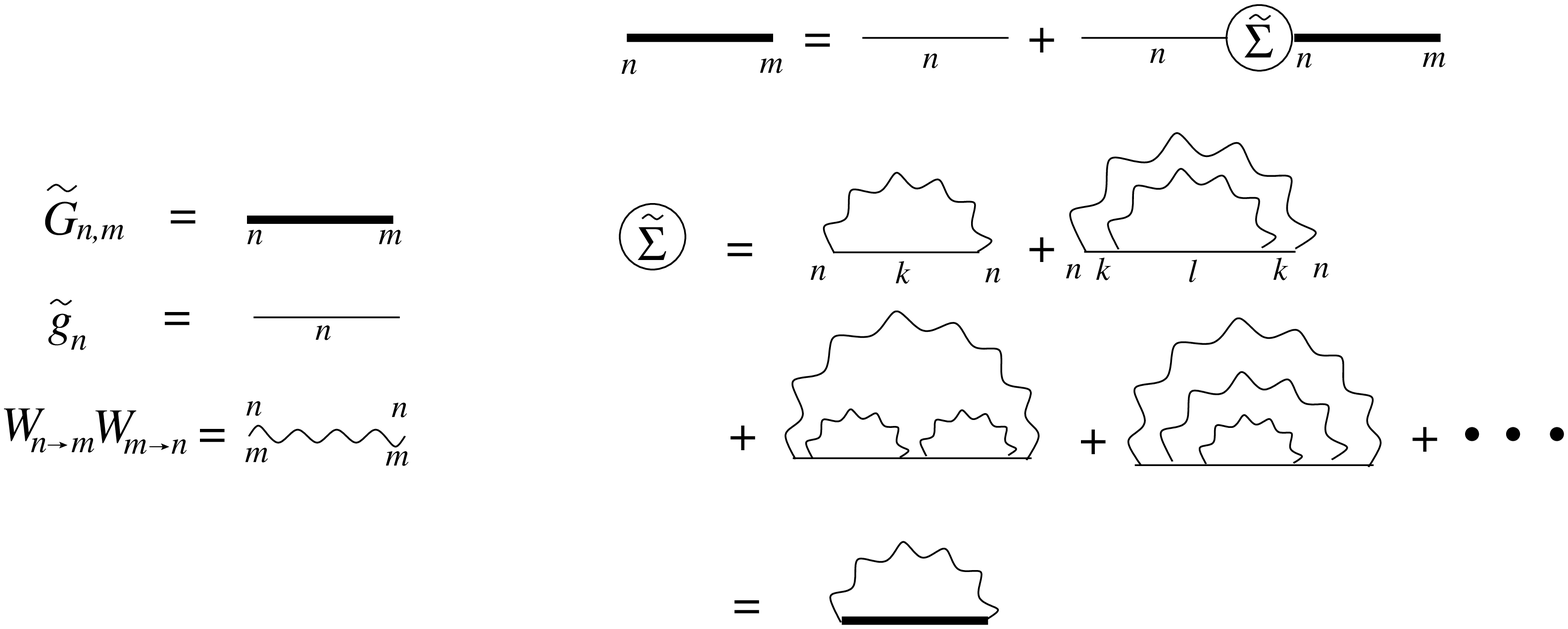}}} \caption[]{Diagramatic expansion of the Green function of the Meiss-ott tree model}
\end{figure}

From the Master equation (\ref{eq:Master}) and from the definition of the Green function (\ref{GFDef}) we obtain:
\begin{eqnarray}
\sum_n \left((s + \mu_n)\delta_{j,n} - \lambda W_{n\to j} \right) \tilde{G}_{n,m}(s) = \delta_{j,m} \label{CGF-eq}
\end{eqnarray}
where
\begin{eqnarray}
\mu_n=\sum_k W_{n\to k}
\end{eqnarray}
and $\lambda$ is a dummy parameter introduced for the expansion of the Green function as a perturbation series, and should be set to one at the end of the calculation.  In particular to the zeroth order in $\lambda$,  $\tilde{G}_{n,m}(s) \approx \tilde{g}_{n}(s) \delta_{n,m}$ where
\begin{eqnarray}
\tilde{g}_{n}(s)=\frac{1}{s+1+\mu_n}.
\end{eqnarray}
Now let us expand the Green function $\tilde{G}_{n,m}(s)$ to all orders in $\lambda$. The diagrammatic representation of this expansion is presented in Fig.~8. Here the thick line represents the exact Green function, thin lines represent the zeroth order Green function, $\tilde{g}_{n,m}(s)$, and wiggly lines represent the product $W_{n \to m}W_{m,\to n}$. Notice that in this expansion wiggly lines cannot cross because there are no loops on Cayley trees. Thus one may write an equation for the Green function $\tilde{G}_{1,1}(s)$ :
\begin{widetext}
\begin{eqnarray}
\tilde{G}_{1,1}(s) = \tilde{g}_{1}(s) + \tilde{g}_{1}(s) \!\sum_{j=0,1} W_{1 \to 1j} \tilde{G}_{1j,1j}^{(1)}(s) W_{1j \to 1 } \tilde{G}_{1,1}(s)
\end{eqnarray}
where the sum over $j$ is the  over the nearest neighbors sites $10$ and $11$, and $\tilde{G}_{1j,1j}^{(1)}(s)$ denotes the Green function on site $1j$ assuming that if the particle reaches site $1$ it disappears. From the self similarity of the Cayley tree it follows that $\tilde{G}_{1j,1j}^{(1)}(s)$ and $\tilde{G}_{1,1}(s)$ are related by a simple rescaling of time:
\begin{eqnarray}
 \tilde{G}_{1j,1j}^{(1)}(s)=\frac{W_{1 \to \phi}}{W_{1j \to 1}} \tilde{G}_{1,1}\left(\frac{W_{1 \to \phi}}{W_{1j \to 1}} s\right) =  \frac{1}{\varepsilon_j} \tilde{G}_{1,1}\left(\frac{s}{\varepsilon_j}\right). \label{AppAEq2}
\end{eqnarray}
Thus the Green function satisfies the equation:
\begin{eqnarray} \left[ s+1+\omega_0+\omega_1- \omega_0 \tilde{G}_{1,1}\left(\frac{s}{\varepsilon_0}\right)- \omega_1 \tilde{G}_{1,1}\left(\frac{s}{\varepsilon_1}\right)\right] \tilde{G}_{1,1}(s) =1 \label{exactG11} \end{eqnarray}
 \end{widetext}
 Now substituting (\ref{tildeexp}) into this equation gives in a straightforward manner equation for $a(0)$,
 \begin{eqnarray}
  \left[ 1+\omega_0+\omega_1- \omega_0 a(0) - \omega_1 a(0)\right] a(0) =1 \end{eqnarray}
 whose solution is
 \begin{eqnarray} a(0)=1. \label{a_0sol}
 \end{eqnarray}
 Now substituting $\tilde{G}_{1,1}(s) = 1+ b(0) s^\gamma +\cdots$ one obtains the equation
 \begin{eqnarray}
 b(0) \left[ s^\gamma - \omega_0 \left(\frac{s}{\varepsilon_0} \right)^\gamma -  \omega_1 \left(\frac{s}{\varepsilon_1} \right)^\gamma\right]= 0
 \end{eqnarray}
 which is equivalent to the dispersion equation  (\ref{eq:gamma1}).

 \section{Appendix B: Disorder diagrammatics}

 In this appendix we compute the correlation function of two Green functions $G_{j,k}(s)$. Our main focus is the disconnected part of the correlator \ref{TGF-correlator}, and we limit our considerations to the the lowest order perturbation theory in $\sigma$, for which the result is independent of the precise distribution of the random variables $\xi_{n,m}$.

 Taking into account the random component of the transition rates we may write the equation for the Green function in the form:
 \begin{eqnarray}
 \sum_m \left(\tilde{G}^{-1}_{n,m}(s)+ \Xi_{n,m}  \right) G_{m,k}(s)= \delta_{n,k}
 \end{eqnarray}
 where
 \begin{eqnarray}
 \Xi_{n,m}=\left( \sum_l W_{n\to l}\xi_{n,l} \right)\delta_{n,m}- W_{m\to n}\xi_{m,n} \label{Xidef} \end{eqnarray}
 is the addition to Eq. (\ref{CGF-eq}) which comes from the random component of the transition rates.

 From the definition of $\xi_{nm}$ (see Eq. (\ref{xi-average})) we have $\langle \Xi_{n,m}\rangle=0$ and
 \begin{widetext}
 \begin{eqnarray}
 \left\langle \Xi_{n,k} \Xi_{l,m }\right\rangle =\sigma^2 \left( \sum_v W_{ n\to v}^2 \delta_{n,k}\delta_{k,m} \delta_{n,l} + W_{n\to l} W_{l \to n} (\delta_{n,k} \delta_{l,m}+ \delta_{l,k} \delta_{n,m})  + W_{k\to n}^2 \delta_{k,m} \delta_{n,l} \right)  \nonumber \\ - \sigma^2 \left( W_{k \to n}^2 \delta_{l,m} \delta_{k,m} + W_{k \to m}W_{m \to k} \delta_{n,l}( \delta_{ n,m}+ \delta_{l,k}) +W_{m \to l}^2 \delta_{n,k} \delta_{k,m}  \right) \label{Wick's}
 \end{eqnarray}
 \end{widetext}
 Notice also that this expression satisfies the relation:
 \begin{eqnarray}
  \sum_n \left\langle \Xi_{n,k} \Xi_{l,m}\right\rangle =0 \label{XiRel}
 \end{eqnarray}
which follows immediately from the definition (\ref{Xidef}).

\begin{figure}[tbh]
\centerline{\resizebox{0.46 \textwidth}{!} {\includegraphics{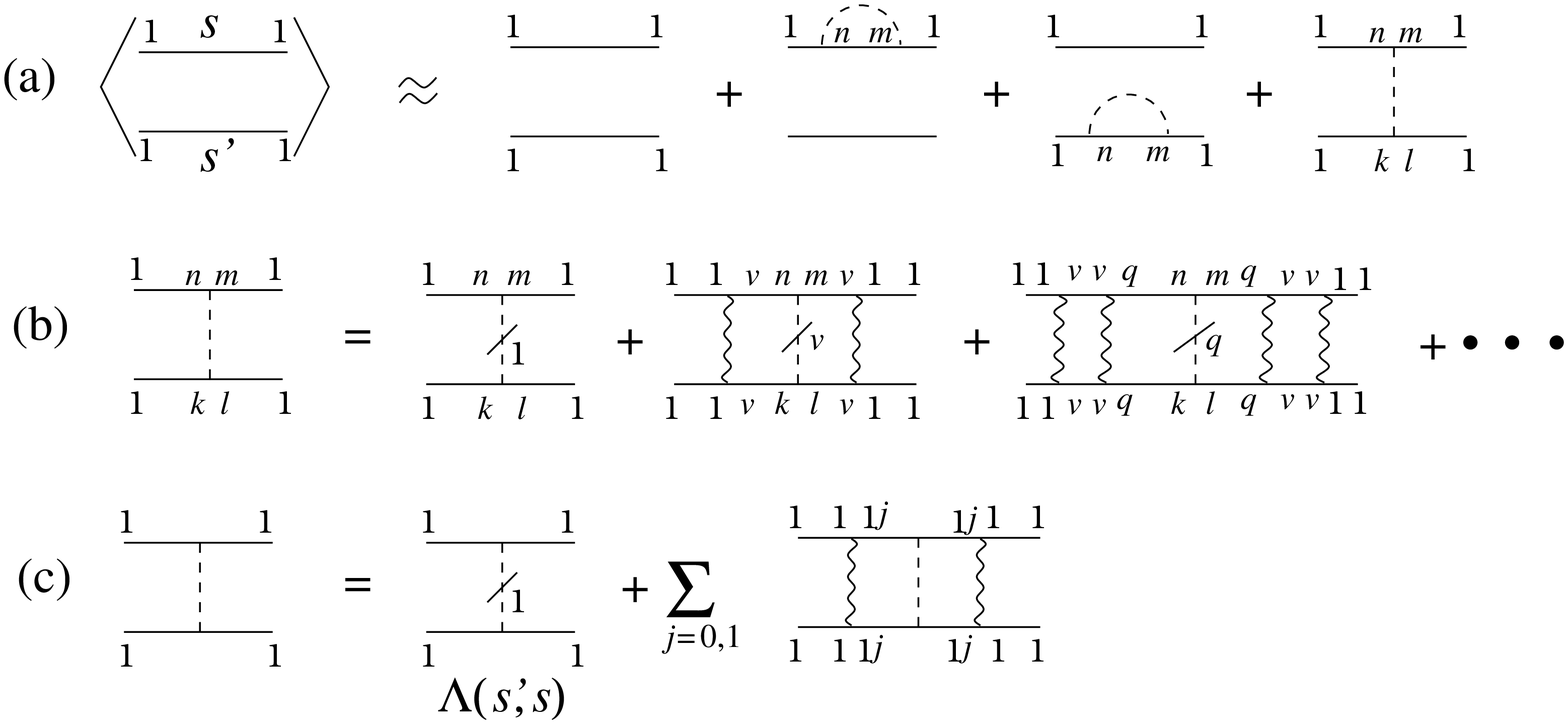}}} \caption[]{Diagrammatic expansion of the correlator  $\langle G_{1,1}(s')G_{1,1}(s)\rangle$.  The dashed line crossed by a segment stands for a sum over all diagrams containing one or more inner indices equal to the index which appears near the segment.}
 \end{figure}

Consider now the disconnected part of two Green functions (\ref{TGF-correlator}). The general diagrams describing the leading order expansion of the average  $\langle G_{1,1}(s')G_{1,1}(s)\rangle$  (up to $\sigma^2$) are shown in the upper panel of Fig.~9.  Here solid lines represent the exact Green function of the system without the random component, while dashed lines stand for pairs of random components of the transition rates given by (\ref{Wick's}).

\begin{figure}[tbh]
\centerline{\resizebox{0.45 \textwidth}{!} {\includegraphics{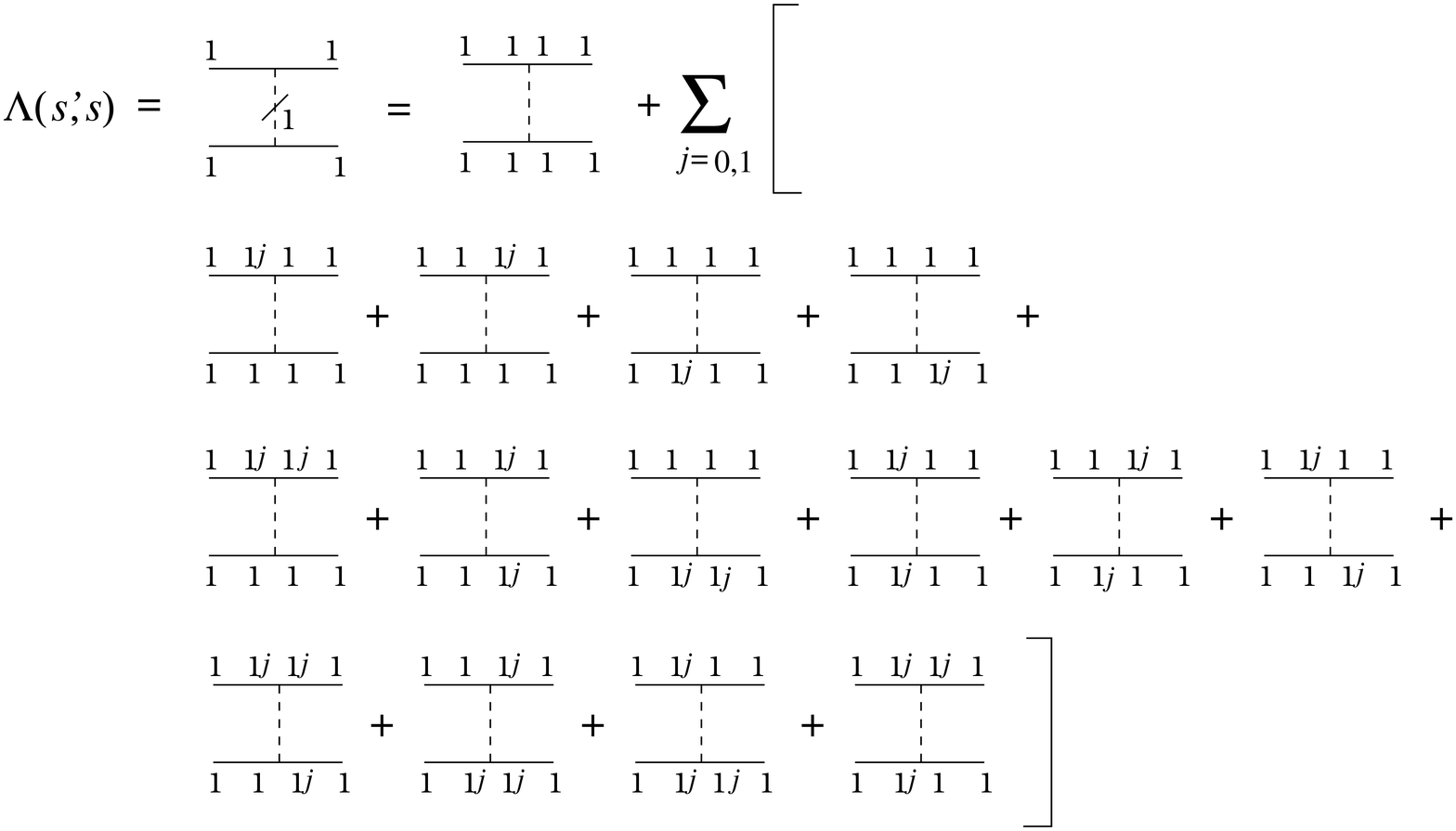}}} \caption[]{The diagram $\Lambda(s's)$}
\end{figure}

The first tree diagrams shown in panel (a) represent the disconnected part of the correlator $\langle G_{1,1}\rangle \langle (s')G_{1,1}(s)\rangle$, which is of no interest here and therefore to the leading order in $\sigma$, the connected part of the correlator
\begin{eqnarray} Q(s's)= \langle \delta G_{1,1}(s')\delta G_{1,1}(s)\rangle \label{defQ}
\end{eqnarray}
is given by the diagram shown in panel (b). The expansion on the right hand side of this panel refers to the point along the tree in which the random contribution is taken into account. The first diagram which will be denoted henceforth by $\Lambda(s's)$ corresponds to the case where the random contribution is taken into account along a bond from site "1" to one of its nearest neighbors sites. The next diagram represents  the case where disorder is taken between site $j$ and its nearest neighbor sites, and so on. Panel (c) of this figure shows the resummation of all these diagrams. From here it follows that
\begin{widetext}
\begin{eqnarray}
\langle \delta G_{1,1}(s')\delta G_{1,1}(s)\rangle = \Lambda(s',s) +\sum_{j=0,1} \tilde{G}_{1,1}(s)^2\tilde{G}_{1,1}(s')^2 W_{1\to 1j}^2 W_{1j\to 1}^2\langle \delta G_{1j,1j}(s')\delta G_{1j,1j}(s)\rangle  \label{resum1}
\end{eqnarray}

Within the leading order in $\sigma$ one may assume that the Green functions satisfy the same scaling relation (\ref{AppAEq2}) and therefore
\begin{eqnarray}
\langle \delta G_{1j,1j}(s')\delta G_{1j,1j}(s)\rangle =  \frac{1}{\varepsilon_j^2} \left\langle \delta G_{1,1}\left(\frac{s'}{\varepsilon_j}\right)\delta G_{1,1}\left(\frac{s}{\varepsilon_j}\right)\right\rangle. \end{eqnarray}
Substituting this relation to (\ref{resum1}) and using definition (\ref{defQ}) one obtains  Eq.~(\ref{Q-eq}).
\end{widetext}

Finally let us calculate the diagram $\Lambda(s',s)$ whose explicit expansion is shown in Fig.~10.  Here solid  lines stand for the exact Green function of the pure system, while dashed lines represent the disorder. Taking into account the relations:
\begin{widetext}
\begin{eqnarray}
\tilde{G}_{1,1j}(s)= \tilde{G}_{1,1}(s)W_{1 \to 1j} \tilde{G}_{1j,1j}^{(1)}(s)= \frac{\omega_j}{\varepsilon_j} \tilde{G}_{1,1}(s)\tilde{G}_{1,1}\left(\frac{s}{\varepsilon_j}\right)  \\ \tilde{G}_{1j,1}(s)= \tilde{G}_{1j,1j}^{(1)}(s) W_{1j \to 1}\tilde{G}_{1,1}(s) = \tilde{G}_{1,1}(s)\tilde{G}_{1,1}\left(\frac{s}{\varepsilon_j}\right)
\end{eqnarray}
\end{widetext}
and using formula (\ref{Wick's}) in order to evaluate the contribution which comes from the the disorder line, leads to Eq. (\ref{LambdaFor}) with
\begin{widetext}
\begin{eqnarray}
\alpha_j(s',s) = \left[ \tilde{G}_{1,1}\left(\frac{s'}{\varepsilon_j}\right) + \tilde{G}_{1,1}\left(\frac{s}{\varepsilon_j}\right) -1 \right]\left[\tilde{G}_{1,1}\left(\frac{s'}{\varepsilon_j}\right) + \tilde{G}_{1,1}\left(\frac{s}{\varepsilon_j}\right) -2\tilde{G}_{1,1}\left(\frac{s'}{\varepsilon_j}\right)\tilde{G}_{1,1}\left(\frac{s}{\varepsilon_j}\right)-1 \right]
\end{eqnarray}
\end{widetext}

\end{document}